# An Interactive Foreign Language Trainer Using Assessment and Feedback Modalities

**Rosalyn P. Reyes**
rosepanelioreyes@gmail.com
**Bulacan State University, Philippines**

**Evelyn C. Samson**
evelynsamson@gmail.com
**Bulacan State University, Philippines**

**Julius G. Garcia**
julius.tim.garcia@gmail.com
**Technological University of the Philippines, Philippines**

*Abstract*. English has long been set as the "universal language." Basically most, if not all countries in the world know how to speak English or at least try to use it in their everyday communications for the purpose of globalizing. This study is designed to help the students learn from one or all of the four most commonly used foreign languages in the field of Information Technology namely; Korean, Mandarin Chinese, Japanese, and Spanish. Composed of a set of words, phrases, and sentences, the program is intended to quickly teach the students in the form of basic, intermediate, and advanced levels. This study has used the Agile model in system development. Functionality, reliability, usability, efficiency, and portability were also considered in determining the level of the system's acceptability in terms of ISO 25010:2011. This interactive foreign language trainer is built to associate fun with learning, to remedy the lack of perseverance by some in learning a new language, and to make learning the user's favorite playtime activity. The study allows the user to interact with the program which provides support for their learning. Moreover, this study reveals that integrating feedback modalities in the training and assessment modules of the software strengthens and enhances the memory in learning the language.

## INTRODUCTION

Communication comes in a way or two. Hence, having his communication skills enhanced is an advantage to anyone working in any field, may it be in the field of Information Technology, Education or Engineering. In the advent of technology, several communication mediums are being utilized such as messaging applications and social networking sites which creates social capital and online intercultural exchange among students for educational purposes (Kubota and Garcia, 2017, p. 105).

However, the requirements to deal with various types of people from different countries, with different cultures and languages raise the problem of language fluency or just being conversant and being able to draw simple conversations for a specific situation if necessary. To be able to attain fluency in a language, a student must be able to dedicate his time and immerse in it.

In order to help students learn the most commonly used foreign languages in the field of Information Technology, namely, Korean, Mandarin Chinese, Japanese and Spanish (Reisenauer, 2016), this study is designed to develop an interactive learning platform and to teach the students in the form of basic, intermediate and advance levels on the mentioned languages. It also uses mnemonics and sample sentences to enhance learning experience and interaction of the user through interactive technology. Moreover, it includes audio prompts for words and phrases in order to aid in the proper pronunciation.

# REVIEW OF RELATED LITERATURE

**Feedback and learning**

The learning process integrates assessment for learning in which it is an approach to classroom assessment (Stobart, 2008). Assessment for learning is primarily used to support the learning process. Providing students with feedback is one of the ways of using assessments and their results as learning approach. In this study, it includes feedback which is provided to individual students taking part in its interactive foreign language trainer as assessment of learning. A vital feature of assessment for learning is feedback. Feedback helps students to gain accurate and deep intuitive understanding about their current position in the learning process and as to how they get to their desired position by providing them with enough and needed information (Stobart, 2008). The desired learning outcomes can be obtained by the students as they receive feedback.

There are different types of written feedback in computer-based assessment. In the study of Shute (2008), the feedback type is based on the following distinctions: (1) the specificity of feedback; (2) the complexity of feedback; and (3) the length of feedback. Knowledge of results (KR) is described as a relatively low complexity type of feedback because it tells only if the answer is correct or not. On the other hand, knowledge of correct response (KCR) is with a higher complexity type of feedback in which it gives the correct answer to every incorrect one. Elaborated feedback (EF) is a more complex type of feedback such as an explanation of the correct answer, solution or reference to study further. Aside from the feedback type is timing feedback. Shute (2008) identifies the timing feedback either immediate or delayed feedback. Immediate feedback is given right after answering every item or question. Delayed feedback can vary in ways such as being delayed until a group or block of items or questions has been completed and being delayed the whole day or even later until the assessment has been completed. There are claims made regarding the effects of immediate and delayed feedback (Mory, 2004). In this study, immediate feedback is used where feedback is provided immediately after answering every item in the training quiz.

Furthermore, there are four levels to achieve feedback which are self, task, process and regulation (Hattie and Timperley, 2007). It is an expansion of the developed model created by Kluger and DeNisi (1996). Feedback at the self-level is focused on the characteristics of the learning of the student. Feedback at the task level is used to correct work and being focused on a surface level of learning such as telling the learner if his answer is correct or not. Feedback at the process level is focused on the process being followed by the student in order to complete the given task or work. Lastly, feedback at the regulation level is said to be in relation the processes or methods in thoughts of the students (Hattie and Gan, 2011). Among the four levels, Hattie and Timperley (2007) favor feedback at the process or feedback at the regulation to use in improving student learning. Hence, feedback at the self-level is said to be an ineffective for learning because it does not give the student any information in order to attain learning goals.

Fabienne van der Kleij (2012) connected the theory of Shute (2008) and the theory of Hattie and Timperley (2007) and created a comprehensive view of providing feedback to a student in different ways. In Figure 1, the content of the feedback refers to the feedback type and level of feedback. KR and KCR only connect to the task level as they both give information only to the learner about his answer. On the other hand, EF is connected to the four levels which means EF is possible to be aimed at all levels. Hence, EF at the self-level is said to be an ineffective strategy to learning. Timing feedback has also an important role to learning as feedback can be provided to the student either right after or immediate or delayed.

There are many other variables which influence the relationship of feedback and learning. According to Stobart (2008), there are three conditions that need to satisfy when making an effective and useful feedback. First condition, the learner needs the feedback. It means that a student needs feedback when his present understanding and the learning goal has a gap or hole in between (Hattie & Timperley, 2007). Second condition, the learner receives the feedback in time to use it and the third condition is that learner is willing and has the ability to use the feedback. However, in a computer-based assessment, not all students exert equal attention to feedback in order to attain learning (Timmers and Veldkamp, 2011).

Paying attention to feedback is being influenced by the correct answer and with more attention for answered items that are incorrect (Timmers & Veldkamp). However, the willingness of the learner to pay attention to the feedback decreases as the length of the test increases. According to Stobart (2008), the difficulty of the test, the length of the assessment and the characteristics of the learner can be used to identify how much attention is being paid by the learner to the feedback being given to him and also it possible effect. It is of great importance that the learner is willing and motivated with his ability to use the feedback, being provided with the needed information and resources to enhance his learning. Otherwise, the learner will not be able to use the feedback in its full potential if the needed information or resources are not available (Stobart, 2008). Furthermore, the feedback should be presented to the learner clearly in order for him to process (Mory, 2004) and use it successfully.

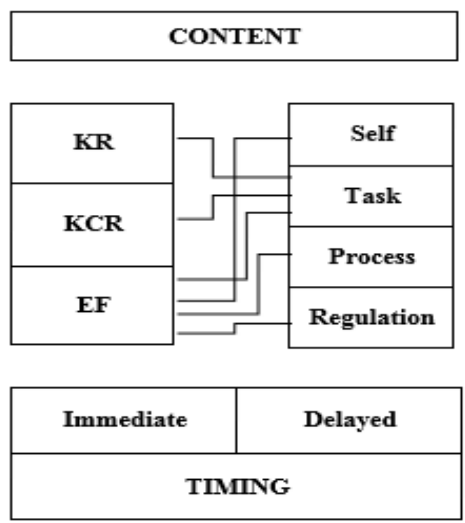

*Figure 1.* A comprehensive view of connecting the theory of Shute (2008) and the theory of Hattie and Timperley (2007)

**Elaborate encoding**

Learning a new one needs to be connected to what a person already knows. Likewise, memories are being made to associate with existing memories. The brain stores facts, ideas or memories, where those which encourage interest to senses, imagination and emotions are said to last longer because of the enjoyment and satisfaction of recalling them.

**Choreographed testing**

The brain muscles are flexible to bend as one recalls his memories. It also makes a memory strong, durable, clear and convincing. There are different ways to strengthen memories such as the more you use your brain to recall a memory, the more you make that memory strong and can easily be recalled.

**Problem Statement**

This study aims to design and developed an interactive language trainer that will help a student learn from one or all the four most commonly used foreign languages in the field of Information Technology; Korean, Mandarin, Japanese and Spanish.

The study also sought to answer the following essential research questions:
1. What is the assessment of students, teachers and IT experts on the performance of the software in terms of functionality, reliability, usability, efficiency and portability? Is there a significant difference among the groups?
2. How does the language training levels utilized to assist and help the students in learning Korean, Mandarin Chinese, Japanese and Spanish?
3. How does the learning assessment and feedback modalities integrated in the software utilized to enhance the learning of the students in Korean, Mandarin Chinese, Japanese and Spanish?

## RESEARCH DESIGN & METHODS

In this study, an evaluation survey was administered to fifty-nine (59) students and twenty-five (25) teachers in the tertiary and twenty-one (21) IT experts in the industry from November to December of 2017 in the Philippines. The survey was conducted to the respondent right after the facilitation and training of the Interactive Foreign Language Trainer application. Table 1 provides the descriptive summary of the respondents.

The data analysis of the developed software is descriptive and quantitative. For the results presentation, analysis and interpretation of data, the following statistical tools are utilized: (1) frequency and percentage distribution in identifying the classification of the respondents, and (2) weighted mean in determining the level of the system's acceptability in terms of ISO 25010:2011 standard characteristics such as functionality, reliability, usability, efficiency and portability. This study adopted a validated scale to develop the survey questionnaire, employing a five-point scale. To facilitate the interpretation of the weighted mean score of the responses, the upper and lower limit of the scale has been adopted using the 5-point Likert Scale, where numerical rating and descriptive interpretation are as follows: 4.60 – 5.00 = Excellent, 3.60 – 4.59 = Very Good, 2.60 – 3.59 = Good, 1.60 – 2.59 = Fair and 1.00 – 1.59 = Poor. One-way ANOVA test was used to identify whether the model is significantly better at predicting the outcome.

Table 1.

*Descriptive Statistics of Respondents*

|  | Frequency | Percent | Mean | SD |
| --- | --- | --- | --- | --- |
| Student | 59 | 56.2 | 51.00 | 17.176 |
| Teacher | 25 | 23.8 | 93.00 | 7.360 |
| IT Expert | 21 | 20.0 | 11.00 | 6.205 |

The study had used the agile method throughout the entire software development process. Agile software development method promotes team collaboration, process adaptability and testing throughout the life cycle. This allows the researchers to ensure that bugs are caught and eliminated in the development cycle, and the software product is double tested again after the previous bug elimination.

Science has been incorporated in designing this software application program. The researchers considered the fact that a person tends to forget what he had just seen in a short period of time, is because a human's short-term memory is the one always activated than the long-term memory. Also, this is what had prompted them to include mnemonics in the program. It takes time to process a short-term memory to a long term one. Mnemonics by definition are devices that help retain information into a form that a human brain can retain better than its original form. Therefore, this study offers foreign language training that combines mnemonics, audio prompt and sequential testing that are believed to strengthen and enhance the memory of its user.

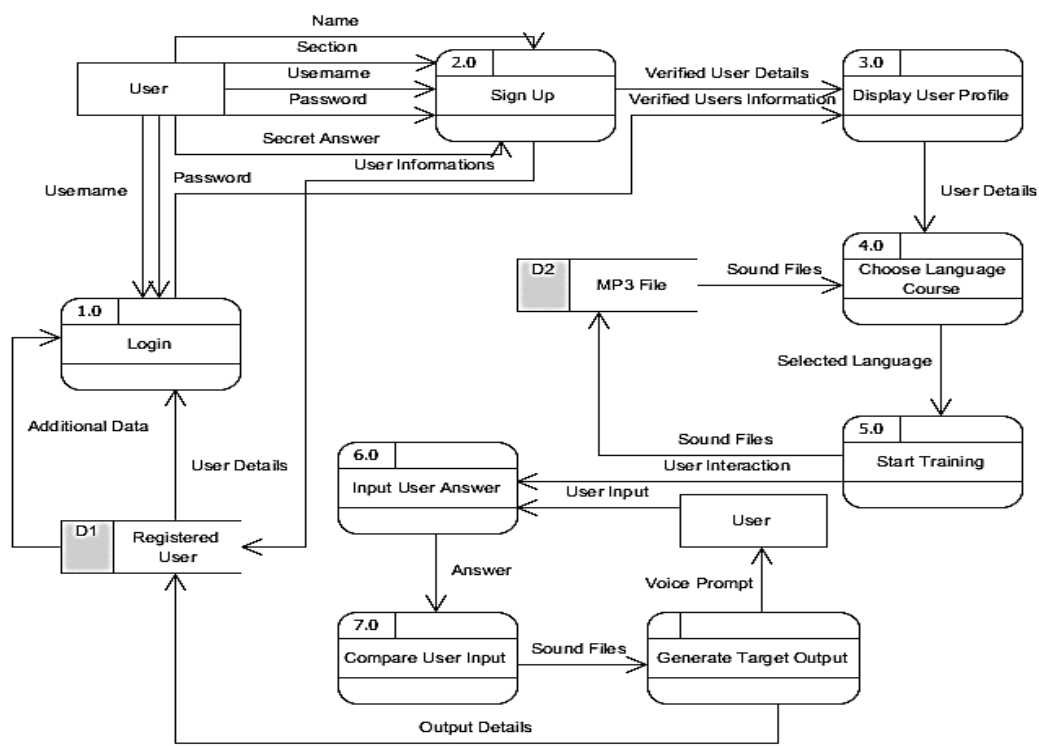

*Figure 2*. Data Flow Diagram of Edify: Interactive Foreign Language Trainer.

The idea of the evaluation or feedback being given to individual user of this interactive trainer is based on Elaborate Encoding and Choreographed Testing in order to provide assessment of learning. Figure 2 shows the data flow diagram to show the different features and functionalities of the developed software which are considered by the researchers necessary to assure learning of the four foreign languages by its individual user.

# RESULTS AND DISCUSSION

The study obtained a total of 105 valid responses from undergraduate students of different year levels, IT Professional/Experts and Teachers (M=26.73, SD=13.96) based on the following: undergraduate students 59 (Male=20, Female=29; M=16.05, SD=1.11) IT Professionals/Experts of 21 (Male=13, Female=8; M=39.72, SD=10.13) and public-school teachers of 25 (Male=4, Female=21, M=41.29, SD=10.96) in the Philippines.

Table 2.

*Descriptive Statistics*

| Criteria | n | Mean | SD | Descriptive Interpretation |
|---|---|---|---|---|
| Functionality | 105 | 4.295 | .618 | Very Good |
| Reliability | 105 | 4.266 | .750 | Very Good |
| Usability | 105 | 4.323 | .713 | Very Good |
| Efficiency | 105 | 4.485 | .637 | Very Good |
| Portability | 105 | 4.142 | .739 | Very Good |

Table 2 contains the five (5) criteria that have been used by the respondents to evaluate the efficacy and acceptability of the developed application.

Table 3.

*Mean Result of groups*

| | Mean | SD | Descriptive Interpretation |
|---|---|---|---|
| Functionality | | | |
| Student (n= 59) | 4.32 | 0.65 | Very Good |
| Teacher (n=25) | 3.95 | 0.56 | Very Good |
| IT Expert (n=21) | 4.30 | 0.62 | Very Good |
| Reliability | | | |
| Student (n= 59) | 4.40 | 0.71 | Very Good |
| Teacher (n=25) | 3.86 | 0.82 | Very Good |
| IT Expert (n=21) | 4.27 | 0.65 | Very Good |
| Usability | | | |
| Student (n= 59) | 4.56 | 0.72 | Very Good |
| Teacher (n=25) | 4.14 | 0.65 | Very Good |
| IT Expert (n=21) | 4.32 | 0.73 | Very Good |
| Efficiency | | | |
| Student (n= 59) | 4.48 | 0.58 | Very Good |
| Teacher (n=25) | 4.10 | 0.51 | Very Good |
| IT Expert (n=21) | 4.49 | 0.77 | Very Good |
| Portability | | | |
| Student (n= 59) | 4.16 | 0.71 | Very Good |
| Teacher (n=25) | 3.95 | 0.80 | Very Good |
| IT Expert (n=21) | 4.14 | 0.74 | Very Good |

For students, the following were the computed mean *(M)* and standard deviation *(SD)* values: system's functionality, M= 4.32, SD= 0.646; system's reliability, M= 4.40, SD= 0.713; system's usability, M= 4.562, SD= 0.720; system's efficiency, M= 4.48, SD= 0.584; and system's portability, M= 4.16, SD= 0.713.  For teachers, the following were the computed mean *(M)* and standard deviation *(SD)* values: system's functionality, M= 3.95, SD= 0.556; system's reliability, M= 3.857, SD= 0.816; system's usability, M= 4.142, SD= 0.650; system's efficiency, M= 4.095, SD= 0.509; and system's portability, M= 3.9524, SD= 0.80.  And for IT Experts, the following were the computed mean (M) and standard deviation (SD) values: system's functionality, M= 4.295, SD= 0.618; system's reliability, M= 4.266, SD= 0.654; system's usability, M= 4.323, SD= 0.727; system's efficiency, M= 4.485, SD= 0.768; and system's portability, M= 4.142, SD= 0.740 respectively.

Overall, the system's functionality (M= 4.295, SD=.618) of the developed system provided functions that were appropriate to the required specifications. Meanwhile, the system's reliability (M= 4.266, SD= .750) of the developed system responded quickly and produced information with consistency. The system's usability (M= 4.32, SD= .713) provided by the developed system was easy to manipulate and understand because of the graphic user interface. System's efficiency has acquired the highest value of M= 4.485 and SD = .637, thus, the developed system provided an accurate information and responded quickly to whatever users' selection. The system's portability (M = 4.12, SD=.739) of the developed system showed that it can be easily installed and can change to new specification or operating environment.

One-way ANOVA was also utilized to analyze and interpret the variance respondents' evaluation results as how useful and acceptable the developed application using ISO 25010:2011 criteria.

Table 4.

*ANOVA Result within and between groups through ISO 25010:2011 standard characteristics*

| Criteria | Mean | SD | df | F | Sig. |
| --- | --- | --- | --- | --- | --- |
| Functionality | 4.2952 | .06041 | 2<br>102<br>104 | 4.480 | .014 |
| Reliability | 4.2667 | .75021 | 2<br>102<br>104 | 4.183 | .018 |
| Usability | 4.3238 | .71381 | 2<br>102<br>104 | 2.164 | .120 |
| Efficiency | 4.4857 | .63722 | 2<br>102<br>104 | 5.906 | .004 |
| Portability | 4.1429 | .73939 | 2<br>102<br>104 | .900 | .410 |

As shown in Table 4, there is significant difference among the 3 groups in terms of functionality, reliability and efficiency of the developed system.

For functionality, there is a significant difference within and between groups, $F(2,102) = 4.480$, p=.014. The students had learned the foreign languages provided in the trainer because it provides the correct result with precision and appropriateness. Likewise, IT experts had seen all the specified tasks and user objectives of the developed system. However, teachers would find the trainer application to be more useful if it allows the user to input words then it will translate to any of the four languages.

For reliability, there is a significant difference between groups and within groups, $F$ (2,102) =4.183, p=.018. The students were able to use features of the trainer under conditions for a period of time. Likewise, IT experts considers the system's performance in terms of maturity, availability, fault tolerance and recoverability. However, teachers would want the trainer application more if the lessons included can be modified or updated as time comes.

For efficiency, there is a significant difference between groups and within groups, $F$ (2,102)=5.906, p=.004. The students had used the trainer application successfully, relative to the amount of resources used under stated conditions. Likewise, IT experts were able to use the trainer correctly, given with time behavior, resource utilization and capacity. However, teachers would want the trainer application more if it may be available to its users using mobile and other similar devices in any time they want.

The differences between the mean values are not likely due to chance and are probably due to evaluation criteria manipulation. Analyzing the data from the ANOVA table, it can be said that IT experts greatly differ to the students and teachers as to how they evaluate the system capabilities and performance. The 3 groups had evaluated the system considering system's navigation, ease of learning, efficiency of use and subjective satisfaction.

**Training levels in the four foreign languages**

Application mode is categorized into two that is integrated and utilized in this study, namely: (1) Training levels in the four foreign languages; and (2) Assessment and feedback.

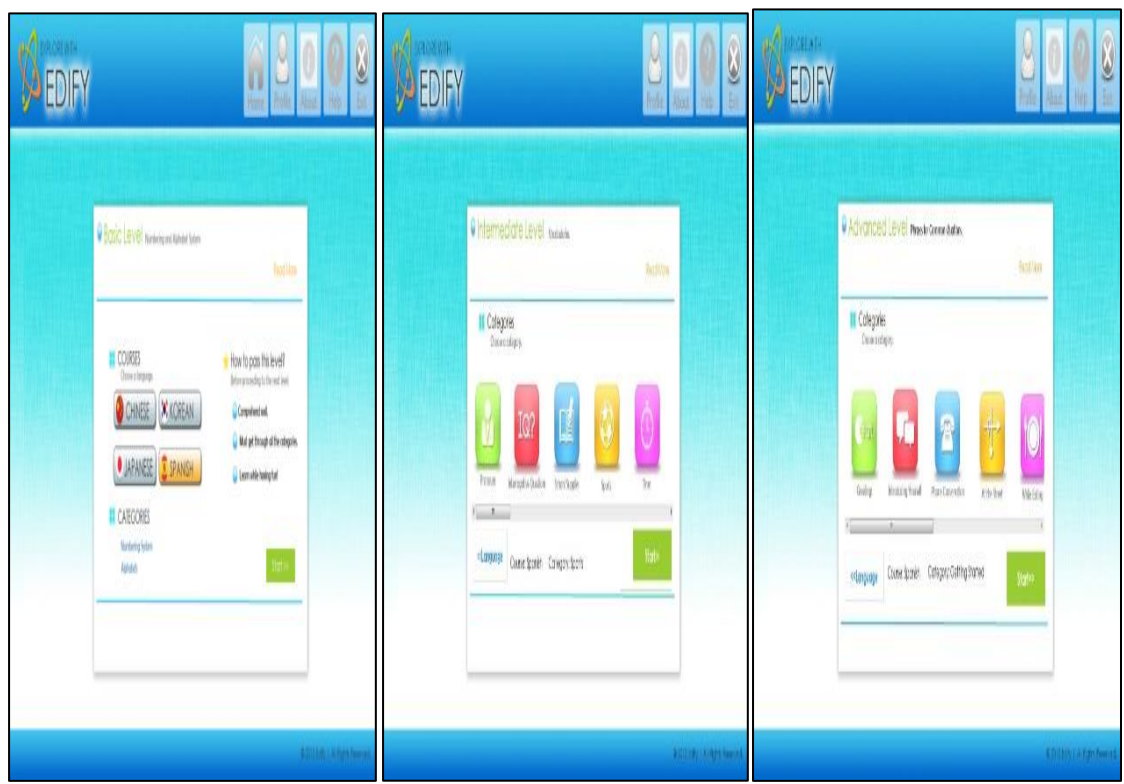

(a) (b) (c)
*Figure 3.* Levels of Training (a) Basic level (b) Intermediate level (c) Advanced level

The three levels of training designed to individual user of the developed interactive foreign language trainer assures that once the user completed the basic level, all the words that have been studied will be shown and the user will be given the option to go to the next level which is intermediate then eventually advanced level. The categories per level serve as tracks of the lessons necessary to achieve the intended learning goals.

In Figure 3 (a), the basic level includes the alphabet and the numbering system of the particular language chosen. In Figure 3 (b), the intermediate level includes pronouns, interrogative questions, school supplies, sports and time reading in any of the four foreign languages. In Figure 3 (c), the advanced level consists of different phrases in a different situation such as phrases being used in greetings, introducing oneself, phone conversation, and wordings at the street and during the time of eating.

**Assessment and feedback modalities**

In Figure 4 (a), the trainer shows the words in English and their translations in other languages with audio. It helps to effectively perform the learning process to individual user as he enjoys the lessons. It combines mnemonics, audio prompt and sequential testing that are believed to strengthen and enhance the memory of its user.

The main purpose of assessment for learning is to support the learning process. Providing the individual user with feedback is one of the ways in using assessments and their results as a learning approach. In this study, it includes feedback which is provided to individual user as they take the quiz. In Figure 4 (b), it tests the user if he has learned anything in the course of his training. After 5 items (words/phrases), 3 questions about the previous items will be given to the user if the quiz toggle is on "ON" mode. It serves as the choreographed testing that aims to strengthen the user's memory.

The type of feedback that was used in this study follows the kind of feedback at the task level where it is mainly intended to correct work and is focused at a surface level of learning. The individual user of the trainer software is informed whether the answer is correct or incorrect by highlighting the indicated answer with color green or red respectively.

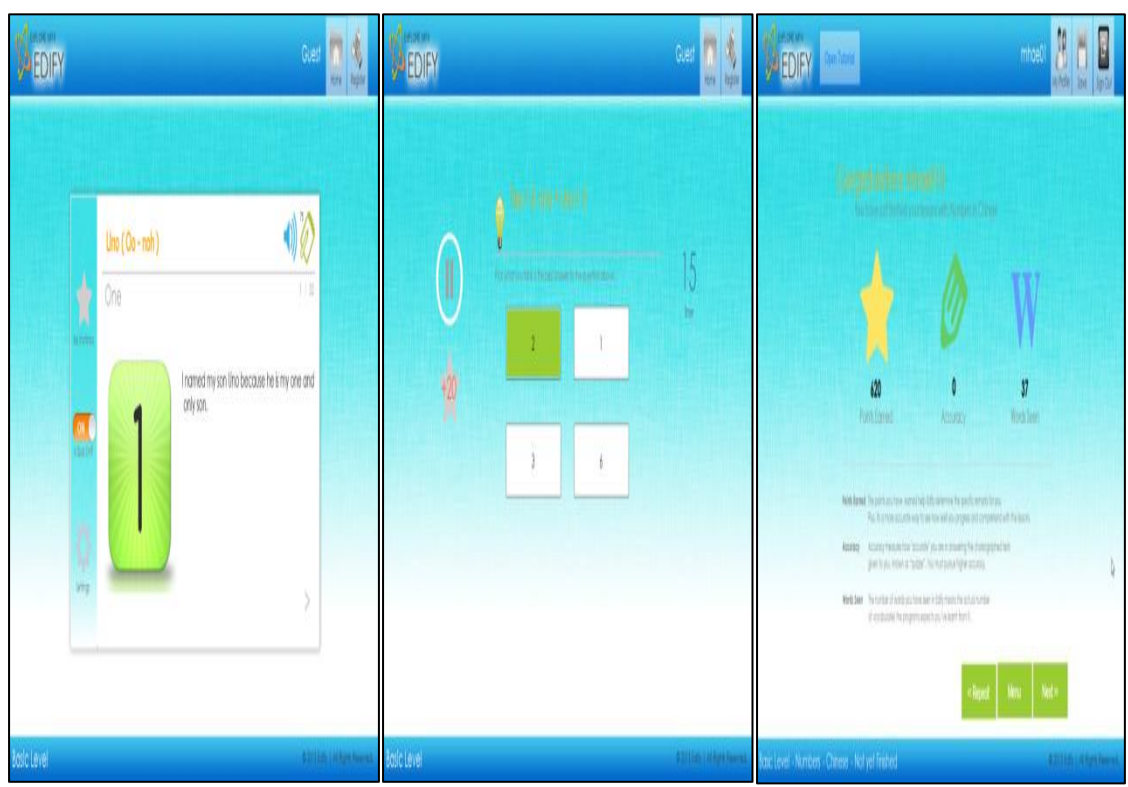

(a)          (b)          (c)

*Figure 4.* Interactive Foreign Language Trainer. (a) The Trainer (b) Quiz and feedback (c) Completion Report per category level

In Figure 4 (c), the completion report contains the assessment of the software application on how the user responded on the training and these are as follows: 1) *Points earned*. The points you have earned helps the Edify determine the specific marks for you. Also, it is a more accurate way to see how well you progress and comprehend with the lessons. 2) *Accuracy*. It measures how accurate you are in answering the choreographed tests given to the user, also known as quizzes. User must pursue higher accuracy. 3) *Words seen*. The number of words the user has seen is the actual number of vocabularies he has learned using the trainer program.

## CONCLUSION

This study aims to design and developed an interactive language trainer that will help a student learn from one or all the four most commonly used foreign languages in the field of Information Technology; Korean, Mandarin, Japanese and Spanish by giving initial training to students. This study also reveals a very good acceptability rating from students, teachers and IT experts. For whatever purpose it may serve, anyone can also use the developed interactive foreign language trainer, may it be used to start training with a foreign language or be used as a course refresher for those who are already familiar with the languages included in the program. It combines mnemonics, audio prompt and sequential testing that are believed to strengthen and enhance the memory of its user. The vocabularies and phrases used by the program to train the users are stored in a database, in such a way it offers a pre-arranged set of words already translated into another language.

Moreover, each language has different levels such as 1) Basic level which includes the alphabet and the numbering system, 2) Intermediate level which includes vocabulary words, and 3) Advance level, it includes the common phrases used for different situations. Assessment of learning is also included by providing feedback to individual user. Every after 5 items (words seen) in each category for every level, there will be 3 questions to be given for the purpose of evaluation and interactivity with user. Specifically, the study used immediate feedback which means that feedback will be given immediately after completion of an item in the training quiz. This will also let the user earn points which will determine the language comprehension level during the training. Lastly, a report is included to assess the user as a newbie, beginner, average or advanced user. However, the application does not include the English language because it is the medium of language in the Tertiary Level and has always been integrated in the curriculum in the Philippines.